\documentclass[epj]{svjour}
\usepackage{graphics}
\usepackage{epsf}

\begin{document}
\renewcommand{\[}{\begin{equation}}
\renewcommand{\]}{\end{equation}}

\title{Superconductivity in striped Hubbard Clusters}
\author{Werner Fettes\inst{1}, Thomas Husslein\inst{1}, 
and Ingo Morgenstern\inst{1}}

\institute{Faculty of Physics, University of Regensburg, 93040 Regensburg, Germany\\
email: Werner.Fettes@physik.uni-regensburg.de}
\date{Received: 16.12.99 / Revised version: date}

\abstract{The CuO-planes of
high-$T_c$ superconductors were found to consist of geometric stripes with
alternating superconducting and antiferromagnetic areas.
Here we will investigate the repulsive Hubbard model
of striped clusters as a possible microscopic description of the superconducting elements.
The focus of our attention lies on the superconducting properties. We report
in agreement with the square Hubbard model a signal in the d$_{x^2-y^2}$-channel and
investigate its dependence on system size, cluster shape and interaction strength.
%%{{\bf Keywords:} High-$T_c$, Superconductivity, Hubbard Model,
%%Off-Diagonal Long Range Order, Stripes, Quantum Monte Carlo}
}  %end of abstract
\maketitle
%

%%\vglue 0.3cm
%%] %%%% Ende von \twocolumn[...

%%%%%%%%%%%%%%%%%%%%%%%%%%%%%%%%%%%%%%%%%%%%%%%%%%%%%%%%%%%%%%%%%%%%%%%%
\section{Introduction}

Short after the discovery of the high-$T_c$ superconductors~\cite{BED86}, the 
Hubbard model was introduced as a generic description of the CuO-planes on 
a microscopic level \cite{AND87}. According to the Van Hove scenario we use 
an extension, the tt'-Hubbard model, to shift the Van Hove singularity in 
the density of states close to the Fermi energy~\cite{NEW92}.
The experimental result of striped domains \cite{CHE89,TSU98}
in the superconducting CuO-planes inside the high-$T_c$ materials has inspired
this study of striped Hubbard clusters.

In order to understand superconductivity in the high-$T_c$ cuprates on a macroscopic level,
the high-$T_c$ glass model was introduced in 1987 \cite{MOR87,MOR89/2}.

It was demonstrated, that the
high-$T_c$ glass model including the tt'-Hubbard model as a microscopic
description of the striped superconducting domains is able to
explain several properties of the high-$T_c$ cuprates \cite{MOR98}, e.g.,~\
the d-wave symmetry of the superconducting phase \cite{WOL93,TSU94}
or the pseudogap above $T_c$ in the density of states \cite{DIN96,LOE96}.
Furthermore this combined high-$T_c$ glass and tt'-Hubbard model picture gives an intuitive
description of the experimental puzzle, that different samples of the same material
and same doping exhibit a nearly constant superconducting transition temperature $T_c$,
yet the critical current densities vary from sample to sample~\cite{MOR99}.

Hubbard clusters were already investigated with numerical algorithms for
a number of different geometries and dimensions, e.g. the one-dimensional chains and
ladders~\cite{DAG92,FAB92,BUL96,NOA96,DAU99}, two-dimensional (2D) 
squares \cite{SAN89,MON94,MOR94,ZHA97/2,FET97/3}, and
layered square systems \cite{BUL92,MOR92,MOR92/3}. 
The stripe instability was found theoretically within Hartree-Fock calculations 
applied to an extended Hubbard model \cite{ZAA89},
and was confirmed by a number of subsequent investigations \cite{ZAA99}.
But up to now it is not known whether the pure Hubbard model exhibits striping, 
e.g., in the form of a phase separation.
In the closely related 2D $t-J$ model the occurrence of stripes is discussed 
controversially \cite{HEL95,HEL98,WHI99,WHI99/2}.
Here we study striped clusters of the Hubbard model directly and do not examine
the occurrence of stripes per se, but only the existence of
superconductivity in striped Hubbard clusters.

In a single striped domain we consider the tt'-Hubbard model, which is described
in real space by \cite{HUB63,GUT63}:
\[
{\cal H} =
{\cal H}_{kin} + {\cal H}_{pot}
\]
with the kinetic
\[
\label{eqhkin}
{\cal H}_{kin} =
\sum_{i,j , \sigma} t_{i,j} (c^{\dagger}_{i,\sigma} c^{}_{j,\sigma}
 +c^{\dagger}_{j,\sigma} c^{}_{i,\sigma})
\]
and the potential part
\[
{\cal H}_{pot} =
U \sum_{i} n_{i,\uparrow} n_{i,\downarrow}
\label{glei1}
\]
of the Hamiltonian.  We denote the creation operator for an electron with
spin $\sigma$ at site $i$ with $c^\dagger_{i,\sigma}$, the corresponding
annihilation operator with $c^{}_{i,\sigma}$, and the number operator at site
$i$ with $n_{i,\sigma} \equiv c^\dagger_{i,\sigma} c^{}_{i,\sigma}$. The hopping
$t_{i,j}$ is only nonzero for nearest neighbors $i,j$ ($t_{i,j}=t$) and
next nearest neighbors ($t_{i,j}=t'$). Finally $U$ is the interaction.
Usually we choose $t'<0$ to shift the Van Hove singularity in the density
of states close to the Fermi energy for less than half filled systems
($\langle n \rangle < 1$, where $\langle n \rangle \equiv (n_{e,\uparrow} + n_{e,\downarrow})/2$
and $n_{e,\sigma}$ is the number of electrons with spin $\sigma$).
Throughout this paper we set $n_{e,\uparrow} = n_{e,\downarrow} \equiv n_e$ and the energy
unit as $t=1$. Additionally we apply periodic boundary conditions both in x- and y-direction.

%%%%%%%%%%%%%%%%%%%%%%%%%%%%%%%%%%%%%%%%%%%%%%%%%%%%%%%%%%%%%%%%%%%%%%%%
\section{Hubbard model and superconducting correlations}

Following \cite{SCA89,FRI91}  we use the (vertex)
correlation function (instead of the largest eigen value of the reduced two particle
density matrix)
as an indicator of superconductivity, making use of the
standard concept of off-diagonal long range order (ODLRO) \cite{YAN62}.

We concentrate here on
the d$_{x^2-y^2}$-wave symmetry (abbreviated as d-wave).
The full two-particle correlation function is defined for the
the d-wave symmetry by
\[
\label{eqfulldxmy}
C_{d}(r) =
\frac{1}{L}
\sum\limits_{i,\delta,\delta'} g_\delta g_{\delta'}
\langle c_{i,\uparrow}^\dagger c_{i + \delta,\downarrow}^\dagger
c_{i+r + \delta',\downarrow}^{} c_{i+r,\uparrow}^{}
\rangle
\quad .
\]
The vertex correlation function $C^V_{d}(r)$ is the two-particle correlation
function $C_{d}(r)$ without the contributions of the single-particle
correlations of the same symmetry \cite{FRI90}.
For the  d-wave the result is
\[
\label{eqvertexdxmy}
C^V_{d}(r) = C_d(r) -
\frac{1}{L} \sum\limits_{i,\delta,\delta'} g_\delta
g_{\delta'}
C_\uparrow (i,r) C_\downarrow (i+\delta,i+r+\delta') \quad .
\]
In equation (\ref{eqvertexdxmy})
$C_\sigma (i,r) \equiv \langle c^\dagger_{i,\sigma} c^{}_{i+r,\sigma}\rangle$
is the single-particle correlation function for spin $\sigma$.
The phase factors are $g_\delta$, $g_{\delta'} =\pm 1$ to model the
d-wave symmetry, the number of lattice points is $L$ and
the sum $\delta$ (resp.\ $\delta'$) is over all nearest neighbors.

We averaged the vertex
correlation function $C^V_d(r)$ only in the large range regime of $r$, i.\,e.\ for the
distances $|r|>|r_c|$:
\[
\bar{C}^{V,P}_d \equiv  \frac{1}{L_c} \sum_{r,|r|>|r_c|} C^V_d(r)
\]
with the number $L_c$ of lattice points with $|r| > |r_c|$.
The qualitative behavior of our results (concerning the vertex correlation
functions) is not influenced by our
choice of $|r_c|$ as long as we suppress the short range correlations
(i.\,e.\ $r_c\ge 1.9$).

Evidence for ODLRO in the d$_{x^2-y^2}$ channel was already found for the case of
the square 2D tt'-Hubbard model \cite{MOR94,HUS94,FET97/3}. We report here the existence of ODLRO in
the striped Hubbard model and investigate the influence of shape and
interaction strength on the superconducting signal.

\begin{figure}[hbtp]
\begin{center}
\begin{minipage}{8.0cm}

\epsfxsize 8.0cm \epsfbox{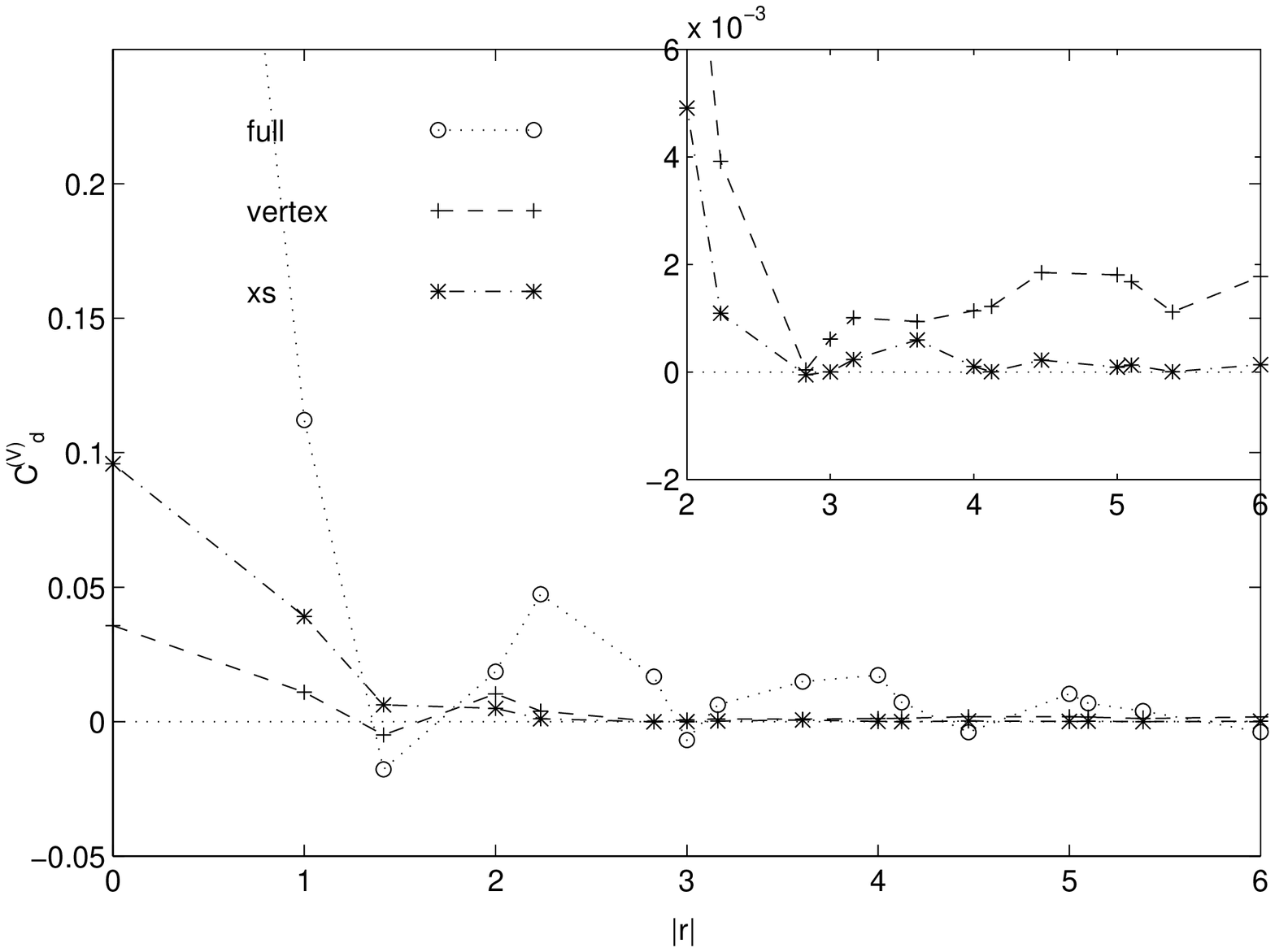}
\end{minipage}
\vglue0.2cm
\caption{\label{fig_long}
Distance $|r|$ between pair creation and pair annihilation operators versus
the two-particle resp.\ vertex correlation function of the Hubbard model 
with $U=2$.
System parameters: $L_x= 12$, $L_y = 4$, $n_e = 21$, $t'=-0.22$, $\theta = 8$, 
$m=64$.
The points labeled with {\em full} and {\em vertex} are correlation functions 
with d-wave symmetry and the points labeled with {\em xs} show the vertex 
correlation function with xs-symmetry.
}
\end{center}
\end{figure}

Figure \ref{fig_long} shows the d-wave correlation functions (eq. (\ref{eqfulldxmy})
and (\ref{eqvertexdxmy})) as a function of the distance $|r|$ between the pair creation and
pair annihilation operators
% $\Delta^{\dagger}_d$ and $\Delta_d$
of both the vertex and the full correlation function for a striped system.
Similar to results for square systems, we obtain huge correlations for the short range part
and finite, positive values for $C^V_{d}(r)$
for large distances in the system (inlay of figure \ref{fig_long}).
Here it becomes
obvious that the vertex correlation function is non-negative in the d-wave case.

For a comparison we plot in figure \ref{fig_long} also the correlation function for the
extended s-wave (xs-) symmetry for the nearest neighbors.
This symmetry obeys the same formulas as equations (\ref{eqfulldxmy}) and (\ref{eqvertexdxmy})
only the phase factors $g_{\delta}$ and $g_{\delta'}$ both are set equal to 1.
In contrast to the d-wave case, the xs-wave symmetry does not exhibit this long range
behavior (inlay of figure \ref{fig_long}). This is also in agreement with simulations for
the square Hubbard model~\cite{MOR94,HUS96}.

%%%%%%%%%%%%%%%%%%%%%%%%%%%%%%%%%%%%%%%%%%%%%%%%%%%%%%%%%%%%%%%%%%%%%%%
\section{The PQMC-Method}

We calculate the ground state properties of the Hubbard model using the projector quantum
Monte Carlo (PQMC) method~\cite{BLA81,KOO82}.
In this algorithm the ground state is projected with
\[
   \vert \Psi_0\rangle = \frac{1}{\cal N} e^{-\theta \cal H} \vert \Psi_T \rangle
\]
from a test state $\vert \Psi_T \rangle$ with the projection parameter $ \theta $ and the normalization factor
$ \cal N $ \cite{LIN92}. In order to perform this projection it is necessary to transform the many-particle
problem into a single-particle problem. This is done in two steps, first the exponential
of the Hamiltonian $ \cal H $ is decomposed into two separate parts, ${\cal H}_{kin}$ and ${\cal H}_{pot}$,
using a Trotter Suzuki transformation \cite{SUZ76,LIN92}
and second the interaction term is treated with a discrete Hubbard Stratonovich (HS) transformation,
which leads to an effective single-particle problem with additional fluctuating HS fields \cite{HIR83}.

We use the second order Trotter Suzuki transformation, which reads as
\[
  e^{-\theta ( {\cal H}_{kin} + {\cal H}_{pot})} =  
  \left( e^{-\frac{\tau}{2} {\cal H}_{kin}}e^{-\tau {\cal H}_{pot}} 
  e^{-\frac{\tau}{2} {\cal H}_{kin}} \right) ^m + {\cal O} (\tau^2) 
\quad ,
\]
where $m$  is the number of Trotter slices and $\tau \equiv \frac{\theta}{m}$. 
Here a systematic error of
order $ {\cal O} (\tau ^2)$ enters the calculations for finite $m$.

The two parameters $m$ and $\theta$ influence the correct projection of the
ground state $|\Psi_0\rangle$ from the test wave function $|\Psi_T\rangle$
in the PQMC algorithm \cite{FET98/3,FET98/4}.

\begin{figure}[hbtp]
\begin{center}
\begin{minipage}{8.0cm}

\epsfxsize 8.0cm \epsfbox{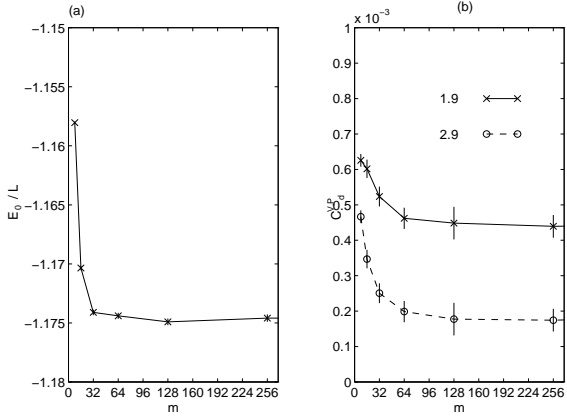}
\end{minipage}
\vglue0.2cm
\caption{\label{fig_m_scal}
$m$-scaling.
System parameters: $L_x= 12$, $L_y = 4$, $n_e = 15$, $U=2$, $t'=-0.22$, $\theta = 8$,
full line $|r_c|=1.9$, dashed line $|r_c|=2.9$.
}
\end{center}
\end{figure}

In figure~\ref{fig_m_scal} we investigate the dependence of the ground state
energy per site $E_0/L$ and of  the vertex correlation function $\bar{C}^{V,P}_d$
on the Trotter parameter $m$. Both, $E_0/L$ and $\bar{C}^{V,P}_d$,
level off for large $m$, indicating the convergence of the PQMC method.
The results resemble similar
PQMC simulations
for the square 2D-tt'-Hubbard model \cite{BOR92,ZHA97,FET98,FET98/4} and 
the APEX-oxygen model \cite{FRI90}.
Like in these cases the vertex
correlation function is here more sensitive to $m$ than the ground state
energy per site $E_0/L$  (figure~\ref{fig_m_scal}).

\begin{figure}[hbtp]
\begin{center}
\begin{minipage}{8.0cm}

\epsfxsize 8.0cm \epsfbox{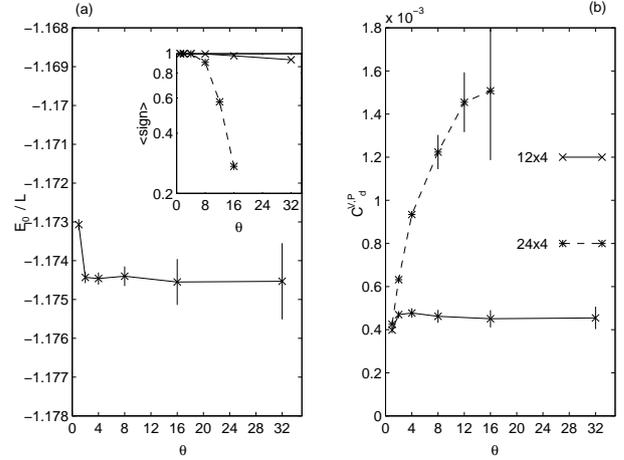}
\end{minipage}
\vglue0.2cm
\caption{\label{fig_theta}
$\theta$-scaling.
System parameters: $L_x= 12$, $L_y = 4$, $n_e = 15$, $U=2$, $t'=-0.22$, $\tau=1/8$
(solid lines),
and $L_x= 24$, $L_y = 4$, $n_e = 41$, $U=2$, $t'=-0.22$, $\tau= 1/8$
(dashed line), ($|r_c|=1.9$).
}
\end{center}
\end{figure}

Figure~\ref{fig_theta} shows the influence of the projection para\-meter
$\theta$ on the same observables. The results are again in good agreement 
with similar simulations for
the square Hubbard model \cite{FET97/3,FET98,FET98/4}.

Due to the more rapid convergence of the ground state energy compared 
to the vertex correlation function the relative changes of $E_0/L$  
of figures~\ref{fig_m_scal} (a) and~\ref{fig_theta} (a) are significantly 
different compared to their (b) counterparts showing
the vertex correlation function. This is also expressed by the very different 
scales of the corresponding y-axes.

Figure~\ref{fig_m_scal} (b) shows additionally to  a $ 12 \times 4 $ system
the results for a twice as large $ 24 \times 4 $ system. Convergence occurs  here at higher
values for $ \theta $, namely $ \theta > 16 $. Due to the sign problem 
(inlay of figure~\ref{fig_theta} (a)) we were not able to perform simulations
for $ \theta > 16 $ . Similar effects occur also for PQMC simulations of the
square Hubbard model~\cite{FET98}.

Quantum Monte Carlo simulations are often plagued with the sign 
problem~\cite{ASS90,LOH90}. The average sign $\langle sign \rangle$ enters the 
calculation for the expectation value $\langle A \rangle$  of an observable $A$ by
\[
\langle A \rangle =\frac{\sum_{\sigma,\sigma'} w(\sigma,\sigma') 
A(\sigma,\sigma')}{\sum_{\sigma,\sigma'} w(\sigma,\sigma')} = 
\frac{\langle A^+\rangle - \langle A^- \rangle }{\langle sign \rangle} \quad.
\label{eq_obs}
\]
Here $ \sigma $ and $ \sigma' $ are configurations of the HS field,
$w(\sigma,\sigma')$ is their weight, and $A(\sigma,\sigma')$ is the expectation value of $A$
for $\sigma$ and $\sigma'$~\cite{HIR83}. Now $w(\sigma,\sigma')$ can have both positive and negative
values, thus when used in a Monte Carlo algorithm for a transition propability one uses the right
hand side of equation (\ref{eq_obs}). $\langle A^+\rangle$ and $ \langle A^- \rangle$
denote the separate averages of HS-configurations $\sigma$ and $\sigma'$ with positive resp.
negative weights $w(\sigma,\sigma')$. Generally speaking QMC simulations are only 
meaningful for $ \langle sign \rangle $ close to 1.

The average sign $\langle sign \rangle$ is known to decrease
for increasing
system size $L$, interaction strength $U$ and projection parameter $\theta$ 
(see the inlay of figure~\ref{fig_m_scal} resp.\ \ref{fig_theta} and~\cite{ASS90,LOH90}).
But a small average sign leads among others to large statistical errors in the Monte Carlo
process and renders the simulation results meaningless.

From the above analysis of the dependence of the ground state energy and the
correlation functions on $ m $ and $ \theta $ we conclude, that
the PQMC
simulations are converged for $U=2$ when  $\theta \geq 8$ in smaller systems 
and $\theta \geq 16$ in larger systems. A ratio 
$\tau = \frac{\theta}{m} = \frac{1}{8} $ of the projection parameter and the 
Trotter parameter was found to be sufficient for a correct decomposition.
Due to the sign problem there is only a small range of the parameters, 
where $\theta $ can be chosen sufficiently large, so that
the investigation of the vertex correlation function and its long range behavior is possible.
This is similar to the case of the square Hubbard model~\cite{FET97/3}.

For smaller systems we performed also some simulations of the Hubbard model
using the stochastic diagonalization~\cite{RAE92,FET97/2}, figure~\ref{fig_var_lx}.
They compare also favorably with their PQMC counterparts, which is an additional
indication that the PQMC performs correctly.

%%%%%%%%%%%%%%%%%%%%%%%%%%%%%%%%%%%%%%%%%%%%%%%%%%%%%%%%%%%%%%%%%%%%%%%%
\section{Superconductivity in stripes}

We now investigate the dependence of the superconducting properties
of the striped Hubbard model on system size, shape and interaction strength.

\begin{figure}[hbtp]
\begin{center}
\begin{minipage}{6.0cm}

\epsfxsize 6.0cm \epsfbox{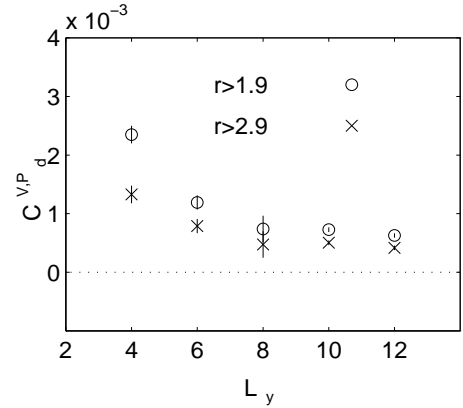}
\end{minipage}
\vglue0.2cm
\caption{\label{fig_var_ly}
Averaged vertex correlation function $\bar{C}^{V,P}_d$ for increasing $L_y$.
System parameters: $L_x = 12$, $U=2$, $t'=-0.22$, $\theta = 8$, $m= 64$, and
fillings: $L_y = 4$: $\langle n\rangle = 0.88$,
$L_y = 6$: $\langle n\rangle = 0.86$,
$L_y = 8$: $\langle n\rangle = 0.77$,
$L_y = 10$: $\langle n\rangle = 0.78$ and
$L_y = 12$: $\langle n\rangle = 0.85$.
}
\end{center}
\end{figure}

\begin{figure}[hbtp]
\begin{center}
\begin{minipage}{5.0cm}

\epsfxsize 5.0cm \epsfbox{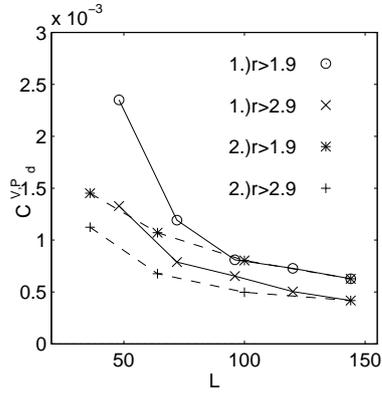}
\end{minipage}
\vglue0.2cm
\caption{\label{fig_size_ly}
Averaged vertex correlation function $\bar{C}^{V,P}_d$ for increasing $L_y$ 
compared  with square system sizes $L = L_x \cdot L_x$.
Simulation parameters: $U=2$, $t'=-0.22$, $\theta = 8$, $m= 64$.
1.) Striped systems: (solid lines), sizes $L_x$, $L_y$, and fillings
$\langle n\rangle$ see figure 4.
2.) Square systems:  (dashed lines),
sizes and fillings: $L_x = 6$: $\langle n\rangle = 0.72$,
$L_x = 8$: $\langle n\rangle = 0.78$,
$L_x = 10$: $\langle n\rangle = 0.82$ and
$L_x = 12$: $\langle n\rangle = 0.85$.
}
\end{center}
\end{figure}

The geometry has a quite significant effect on the magnitude of the correlation functions.
For increasing width $L_y$ of the stripes, figure~\ref{fig_var_ly}, the average long range
part of the vertex correlation function $\bar{C}^{V,P}_d$ is decreasing significantly.
The ratio between $\bar{C}^{V,P}_d$ for a rectangular $ 12\times4$  and a square 
$12 \times 12$ system is almost 3.
In figure \ref{fig_size_ly} we show $\bar{C}^{V,P}_d$ for both,
square systems and the rectangular $12 \times L_y $ systems from figure~\ref{fig_var_ly},
as a function of the system size $L=L_x\times L_y$.
Here the rectangular shaped systems always show a higher superconducting signal than
square systems of the same size $L$.

This is rather surprising when one takes into account that in striped systems, on average, the
distances $|r|$ between pair creation and pair annihilation operators are larger than in
square systems with the same number of sites $L$. In our view there are two effects
which may increase the superconducting correlations. First the anisotropy of
$L_x$ and $L_y$ which leads to more finite size shells. Finite size shells refer to the
energy levels of the free Hubbard clusters ($U=0$).
It is known, that other
ways of introducing additional shells, e.g., anisotropic hopping $t_x \neq t_y$
\cite{KUR96,KUR97}, or
additional hopping to next nearest neighbors~\cite{HUS96} increase the
superconducting correlations in the repulsive Hubbard model.

In our view it is a second effect, the squeezing of the system in one dimension, 
that gives rise to these increased superconducting correlations.

In contrast to the width $L_y$, the length $L_x$ of the stripes is relatively 
insensitive to the height of the plateau, figure~\ref{fig_var_lx}.

\begin{figure}[hbtp]
\begin{center}
\begin{minipage}{8.0cm}

\epsfxsize 8.0cm \epsfbox{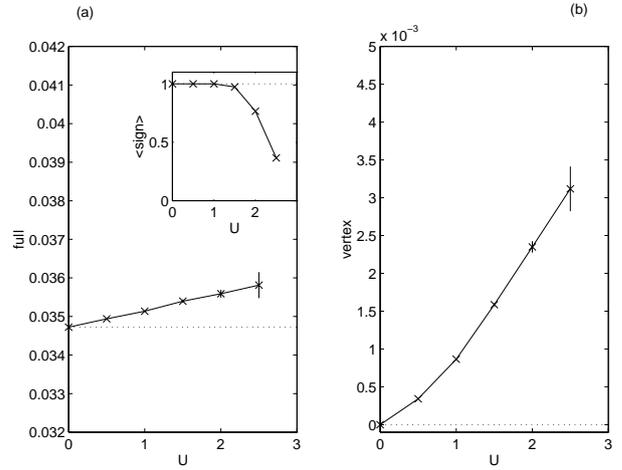}
\end{minipage}
\vglue0.2cm
\caption{\label{fig_var_u}
Averaged  vertex ($|r_c|>1.9$) and full (all lattice points) d-wave correlation
function  for increasing interaction $U$.
System parameters: $L = 12\times 4$, $\langle n\rangle = 0.88$, $t'=-0.22$, 
$\theta = 8$,  $m= 64$.
The inlay shows the average sign.
}
\end{center}
\end{figure}

Another way to strengthen the superconducting correlations is to 
increase the (repulsive) Hubbard interaction U, as shown in figure~\ref{fig_var_u}. 
Here we present both common methods for analyzing superconductivity: the 
full correlation function~\cite{IMA91,SCA89,WHI89} and the vertex 
function~\cite{SCA89,FRI91,FET97/2}. The dotted lines in figure~\ref{fig_var_u}(a) 
and (b) indicate the values of the full (vertex) correlation function in the case 
of no interaction $U=0$. Figure \ref{fig_var_u}(a)
shows that the full correlation function increases for higher interactions.
The vertex correlation function is zero for the case of no interaction ($U=0$) and
increases also monotonous for increased interaction strength.
Due to the sign problem we were not able to perform simulations for an
interaction strength $ U > 2.5 $, for the system size and filling shown in 
figure~\ref{fig_var_u}.

Thus within the range of parameters accessible by the PQMC method, the 
superconducting correlations are increasing for an increasing repulsive 
interaction strength $U$. Both full and vertex correlation
function show this behavior. We want to note, that the vertex correlation 
function is much more sensitive to variations of $U$ due to the substraction 
of the background of the single-particle correlation functions (figure~\ref{fig_var_u}). 
Thus the vertex correlation function is the
more appropriate observable to analyze superconductivity in small Hubbard clusters.
These results are  similar to observations made for the BCS reduced Hubbard model~\cite{FET97/2}.

%%%%%%%%%%%%%%%%%%%%%%%%%%%%%%%%%%%%%%%%%%%%%%%%%%%%%%%%%%%%%%%%%%%%%%%%
\section{Effective interaction in striped Hubbard clusters}

Due to the failure of the usual finite size scaling in the square 2D
Hubbard model we introduced \cite{HUS96,FET97/3,FET98} an effective
model, the BCS-reduced Hubbard model, to compare the superconducting
correlations for different system sizes $L$. This failure is mainly caused by the
underlying shell structure of the free ($U=0$) system \cite{FET98,FET98/3,HUS96}.

The BCS-reduced Hubbard model exhibits the same corrections to scaling as the Hubbard model,
and has a well chosen interaction term, that produces superconductivity with d-wave symmetry.
We calculate for this model the same correlation functions as for the Hubbard model. The effective interaction
strength $J_{eff}$ is then chosen to give the same values for the correlation functions as the
Hubbard model (for details see~\cite{FET98}). From this we get a direct estimate of
the superconducting interaction strength.

The calculation of an effective interaction for the three band Hubbard model
was used to identify the pairing mechanism for d-wave superconductivity in this model.
The evidence of d-wave pairing in this case is based on symmetry arguments and exact
diagonalization results of small clusters
\cite{CIN97,CIN98,CIN98/2,CIN99}.

In the momentum space the BCS-reduced Hubbard model is described by the
Hamiltonian:
\[
\label{eqhbcs}
{\cal H}^{BCS} =  {\cal H}^{BCS}_{kin} +  {\cal H}^{BCS}_{int}
\quad .
\]
The kinetic part is again equation~(\ref{eqhkin}), only transformed to momentum space, i.\,e.\
${\cal H}^{BCS}_{kin} = {\cal H}_{kin} $,
and the interaction is defined by (for d-wave interaction)
\[
{\cal H}_{int}^{BCS} = \frac{J}{L} \sum_{{k,p \atop k \ne p}} f^{}_k f^{}_p
c_{k,\uparrow}^\dagger
c_{-k,\downarrow}^\dagger c_{-p,\downarrow}^{} c_{p,\uparrow}^{}
\quad .
\label{eqhbcsintk}
\]
In equation (\ref{eqhbcsintk}) we use the form factors
\[
\label{eqformd_wave}
f_k \equiv \cos(k_x) - \cos(k_y)
\]
to model the d-wave symmetry in 2D ($k\equiv (k_x,k_y)$).

We calculate the ground state of this BCS-reduced Hubbard model with
the exact and the stochastic diagonalization \cite{RAE92,RAE92/2,FET97/2}.

\begin{figure}[hbtp]
\begin{center}
\begin{minipage}{6.0cm}

\epsfxsize 6.0cm \epsfbox{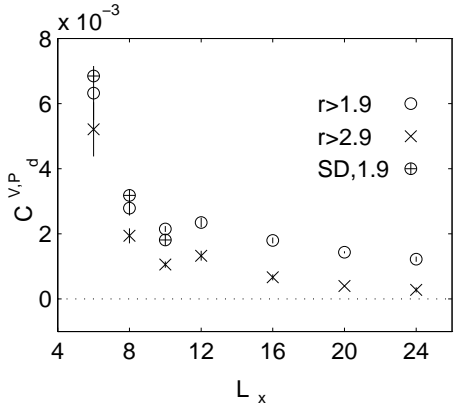}
\end{minipage}
\vglue0.2cm
\caption{\label{fig_var_lx}
Averaged vertex correlation function $\bar{C}^{V,P}_d$ for increasing $L_x$.
System parameters: $L_y = 4$, $U=2$, $t'=-0.22$, $\theta = 8$, $m= 64$.
PQMC runs are averaged for $|r| > |r_c| = 1.9$ (o) and 
$|r| > |r_c| = 2.9$ ($\times$).
SD runs are averaged for $|r| > |r_c| = 1.9$ ($\oplus$).
Sizes and fillings: $L_x = 6$: $\langle n\rangle = 0.92$,
$L_x = 8$: $\langle n\rangle = 0.81$,
$L_x = 10$: $\langle n\rangle = 0.75$,
$L_x = 12$: $\langle n\rangle = 0.88$,
$L_x = 16$: $\langle n\rangle = 0.78$,
$L_x = 20$: $\langle n\rangle = 0.83$ and
$L_x = 24$: $\langle n\rangle = 0.85$.
}
\end{center}
\end{figure}

\begin{figure}[hbtp]
\begin{center}
\begin{minipage}{6.0cm}

\epsfxsize 6.0cm \epsfbox{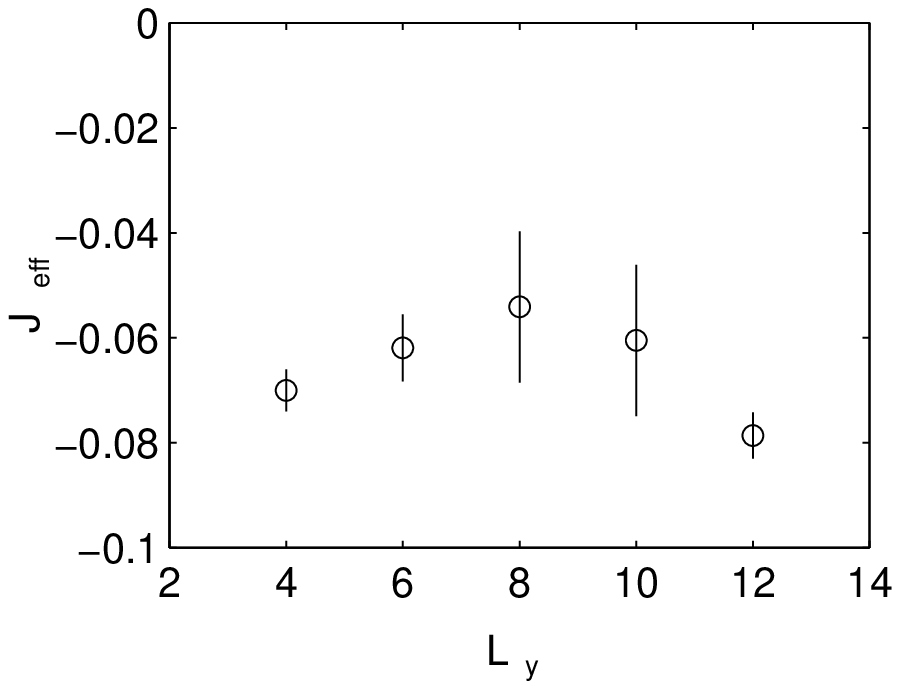}
\end{minipage}
\vglue0.2cm
\caption{\label{fig_j_eff_lx}
Effective d-wave interaction $J_{eff}$ for various $L_y$.
System parameters: $L_x = 12$, $U=2$, $t'=-0.22$, $\theta = 8$,
and $m= 64$.
Fillings see figure 4.
}
\end{center}
\end{figure}

\begin{figure}[hbtp]
\begin{center}
\begin{minipage}{6.0cm}

\epsfxsize 6.0cm \epsfbox{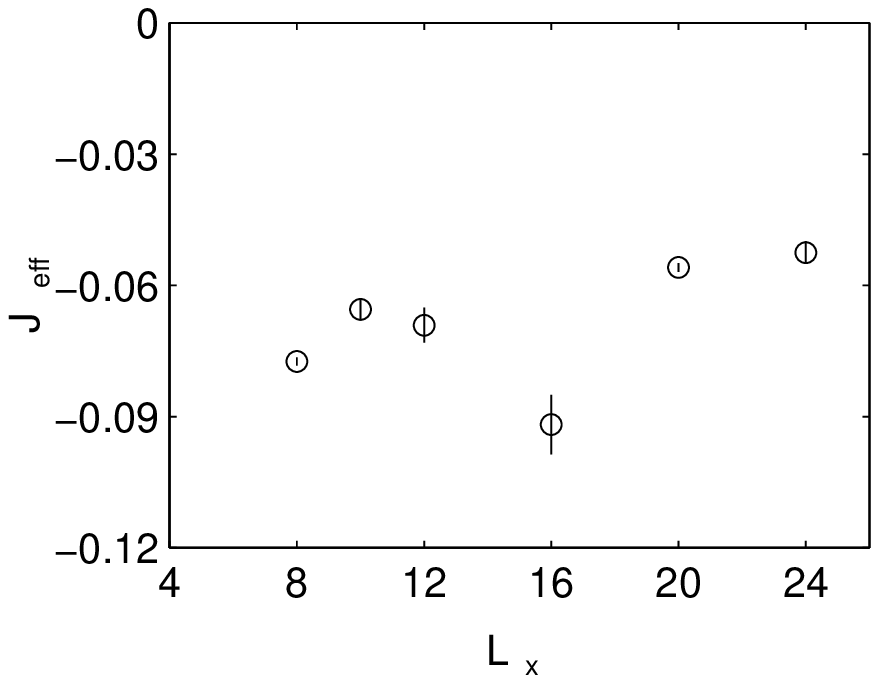}
\end{minipage}
\vglue0.2cm
\caption{\label{fig_jeff_ly}
Effective d-wave interaction $J_{eff}$ for various $L_x$.
System parameters: $L_y = 4$, $U=2$, $t'=-0.22$, $\theta = 8$,
and $m= 64$.
Fillings see figure 6.
}
\end{center}
\end{figure}

In figure \ref{fig_j_eff_lx} and \ref{fig_jeff_ly} we show the effective 
interaction $ J_{eff} $ corresponding to the correlation functions and systems 
shown in figure \ref{fig_var_ly} and \ref{fig_var_lx}.
Within the error bounds of the simulations we conclude that $ J_{eff} $ is nearly constant for
various geometries of the system.
It is not possible  to calculate stochastic errors of the physical
observables within the SD. But for smaller system sizes our comparison of SD
with exact diagonalization
results indicates that for the weak interactions $J$ used here the errors in the SD
are negligible \cite{HUS96,FET97/2}. The error bars shown in figures~\ref{fig_j_eff_lx} 
and \ref{fig_jeff_ly} are therefore calculated using only the statistical errors of 
the PQMC results and fitting these values to the SD results.

In addition to the above mentioned, one has to take into account,
that even so we tried to perform the calculations at a constant 
filling $ \langle n \rangle \approx 0.8 $ the constraint of closed shells for PQMC 
simulations leads to  different fillings $\langle n \rangle$ for
each of these system sizes. Furthermore, in the case of figure \ref{fig_jeff_ly} (and \ref{fig_var_lx} respectively) one has
to take into account, that all simulations are performed at a constant $\theta = 8 $.
Whereas figure \ref{fig_theta} indicates that for large system sizes $L$ a higher value of
$ \theta $ would lead to slightly higher values of the vertex correlation function in the PQMC runs
and thus to a slightly lower effective interactions $J_{eff}$.

From figures \ref{fig_j_eff_lx} and \ref{fig_jeff_ly} we conclude that within the accuracy of
the applied methods, the effective interaction strength $J_{eff}$ is equal for both
square and striped systems. Furthermore $J_{eff}$ is insensitive to the length of the striped systems.

%%%%%%%%%%%%%%%%%%%%%%%%%%%%%%%%%%%%%%%%%%%%%%%%%%%%%%%%%%%%%%%%%%%%%%%%
\section{Summary and Conclusions}

Here we performed ground state simulations of the 2D tt'-Hubbard model and 
the BCS-reduced Hubbard model of striped clusters using PQMC and SD techniques. 
Together with the exact diagonalization these  are the most reliable computational tools
for this type of calculations.

We concentrated our investigations on the behavior of rectangular striped systems.
In agreement with previous calculations of the square Hubbard model we find that
these finite systems show evidence for superconductivity in the $d_{x^2-y^2}$ 
channel for repulsive interactions $U$.
Compared to the squared case these correlations are significantly enhanced, and the superconducting
signal is nearly insensitive to the length of these stripes.

Using SD-techniques we were capable of estimating the effective superconducting interaction
strength $ J_{eff} $ of a BCS-reduced Hubbard model with the same symmetry of the 
superconducting correlation functions. Within the accuracy of our
methods both square and striped Hubbard model show approximately the same 
superconducting interaction strength $ J_{eff} $.

In conclusion, the striped Hubbard model is a  promising candidate for the microscopic
description of the superconducting striped domains in the high-$T_c$ cuprates.
Within the larger framework of the high-$T_c$ glass
model a combined model is able to explain many puzzling properties of the high-T$_c$ materials.

%%%%%%%%%%%%%%%%%%%%%%%%%%%%%%%%%%%%%%%%%%%%%%%%%%%%%%%%%%%%%%%%%%%%%%%%
\section{Acknowledgment}

We want to thank P.C. Pattnaik, D.M. Newns, C.C. Tsuei, T. Doderer, H. Keller, 
T. Schneider, J.G. Bednorz, and K.A. M{\"u}ller for very helpful discussions.
The LRZ Munich grants us a generous amount of CPU time on their IBM SP2 parallel computer,
which is highly appreciated.
Finally we acknowledge the financial support of the UniOpt GmbH, Regensburg.

%%%%%%%%%%%%%%%%%%%%%%%%%%% BIBLIOGRAPHY %%%%%%%%%%%%%%%%%%%%%%%%%%%%%%%
%%%%%%%%%%%%%%%%%%%%%%%%%%%%%%%%%%%%%%%%%%%%%%%%%%%%%%%%%%%%%%%%%%%%%%%%

\bibliographystyle{unsrt}

%%%%%%%%%%%%%%%%%%%%%%%%%%%%%%%%%%%%%%%%%%%%%%%%%%%%%%%%%%%%%%%%%%%%%%%%
%%%%%%%%%%%%%%%%%%%%%%%%%%% F I G U R E S %%%%%%%%%%%%%%%%%%%%%%%%%%%%%%
%%%%%%%%%%%%%%%%%%%%%%%%%%%%%%%%%%%%%%%%%%%%%%%%%%%%%%%%%%%%%%%%%%%%%%%%

\end{document}